\documentclass[5p, times]{elsarticle}

\usepackage{amsmath,amsfonts,amscd,amssymb}
\usepackage{graphicx}
\usepackage{epstopdf}
\usepackage{overpic}
\usepackage{cancel}
\usepackage{rotating}
\usepackage{url}
\usepackage{caption}
\usepackage{color}
\usepackage{rotating}
\usepackage{multirow}
\usepackage{wrapfig}
\usepackage{mathtools}
\usepackage{subeqnarray}
\usepackage{hyperref}
\usepackage{xcolor}

\graphicspath{{Figures/}}

\begin{document}
\begin{frontmatter}
\title{Stochastically forced ensemble dynamic mode decomposition for forecasting and analysis of near-periodic systems}


\author{Daniel Dylewsky\corref{cor1}\fnref{fn1}}
\ead{dylewsky@uw.edu}
\cortext[cor1]{Corresponding author}

\author{David Barajas-Solano\fnref{fn2}}

\author{Tong Ma\fnref{fn2}}

\author{Alexandre M. Tartakovsky\fnref{fn3}}

\author{J. Nathan Kutz\fnref{fn1}}

\fntext[fn1]{University of Washington Department of Applied Mathematics, 
Lewis Hall 201, Box 353925 
Seattle, WA 98195-3925}
\fntext[fn2]{Pacific Northwest National Laboratory}
\fntext[fn3]{University of Illinois at Urbana-Champaign}


\begin{abstract}
Time series forecasting remains a central challenge problem in almost all scientific disciplines. We introduce a novel load forecasting method in which observed dynamics are modeled as a forced linear system using Dynamic Mode Decomposition (DMD) in time delay coordinates. Central to this approach is the insight that grid load, like many observables on complex real-world systems, has an "almost-periodic" character, i.e., a continuous Fourier spectrum punctuated by dominant peaks, which capture regular (e.g., daily or weekly) recurrences in the dynamics. The forecasting method presented takes advantage of this property by (i) regressing to a deterministic linear model whose eigenspectrum maps onto those peaks, and (ii) simultaneously learning a stochastic {\em Gaussian process regression} (GPR) process to actuate this system. Our forecasting algorithm is compared against state-of-the-art forecasting techniques not using additional explanatory variables and is shown to produce superior performance. Moreover, its use of linear intrinsic dynamics offers a number of desirable properties in terms of interpretability and parsimony. Results are presented for a test case using load data from an electrical grid. Load forecasting is an essential challenge in power systems engineering, with major implications for real-time control, pricing, maintenance, and security decisions.
\end{abstract}

\begin{keyword}
Dynamic mode decomposition, data-driven forecasting, Gaussian process regression, power grid, load forecasting
\end{keyword}

\end{frontmatter}

\section{Introduction}
Despite its long history of innovations and mathematical maturity, the analysis of time series data remains an exceptionally active area of research~\cite{box2011time,das1994time,percival2000wavelet,box2015time,shumway2017time,montgomery2015introduction}. Although there can be a diversity of objectives in such analysis, accurate and robust {\em forecasting} is typically the most difficult goal to achieve. Forecasting is fundamentally an {\em extrapolation} problem:  Given a set of measurements over a prescribed time period (training data), predict the future state of the system (extrapolatory test data).
Existing  forecasting techniques are typically classified into two categories:  {\em statistical techniques} (multiple linear regression (MLR) models \cite{hong2010short,hong2013long, charlton2014refined, wang2016electric,xie2016relative,xie2016temperature, xie2017variable}, semi-parametric additive models \cite{hyndman2009density,fan2011short, goude2013local,nedellec2014gefcom2012}, autoregressive integrated moving average (ARIMA) models \cite{boroojeni2015optimal,boroojeni2017novel}, and exponential smoothing models \cite{taylor2007short,taylor2008evaluation}, among others) and {\em machine learning/artificial intelligence} (ML/AI) techniques (e.g., neural networks \cite{yun2008rbf,qingle2010very, shi2017deep, ryu2017deep, bianchi2017overview,zheng2017short,din2017short,lange2020fourier}, fuzzy regression models \cite{song2005short,hong2014fuzzy}, support vector machines (SVMs) \cite{chen2004load,chen2017short}, and gradient boosting machines \cite{taieb2014gradient,lloyd2014gefcom2012}).
Increasingly, the boundary between these categories is blurred, as many ML/AI techniques are simply large-scale, computational statistics algorithms. Integrating methods from both statistics and ML is leading to rich, multidisciplinary collaborations in the scientific community, including for load forecasting in power grid networks. In this article, we integrate the data-driven regression of {\em dynamic mode decomposition} (DMD) with the classic statistical method of {\em Gaussian process regression} (GPR) to develop a robust forecasting tool that improves power grid load forecasting over current, state-of-the-art statistical and ML/AI methods. 

Load forecasting in power grid networks has been of fundamental importance since the inception of the electric power industry~\cite{hong2016probabilistic}.
Many critical decisions, including system maintenance, security and reliability analysis, price and income elasticities, and generation scheduling, are made from load forecasting~\cite{amjady2001short}.
In the past decade, the increased market competition, aging infrastructure, and renewable energy integration requirements has made load forecasting not only more important, but also significantly more difficult~\cite{hong2016probabilistic}. 
This paper presents a purely data-driven method for load forecasting implemented on time series data provided by the Duke Energy power company. Load data from the states of North Carolina and South Carolina is collected every hour over a 365 day period, spanning from 1/1/2017 to 12/30/2017. Load fluctuations are driven by many drivers such as varying weather and seasonal conditions, holidays, work cycles, and fluctuations of an economic nature, which yield data with nuanced temporal patterns.
This makes accurate and robust forecasting of load data very challenging. 

The method introduced here seeks to improve on existing methods by modeling grid load as a forced linear system via {\em Dynamic Mode Decomposition} (DMD) in time delay coordinates. This approach offers a more interpretable algorithmic structure than many state-of-the-art techniques: its linear intrinsic dynamics enable spectral analysis through eigenvalue decomposition. Furthermore, the forcing signal which it discovers can offer diagnostic insight into anomalous events with signatures in the original data. In comparison with other leading methods, we show that for the Duke load data our method consistently outperforms on forecasts out to a horizon of 2 weeks. For the sake of comparison we present results for algorithms which do not make use of additional explanatory variables such as temperature or holidays, which are known correlates to power consumption. Many existing methods have been augmented with such variables to considerable success~\cite{hong2010short,hong2013long, charlton2014refined, wang2016electric,xie2016relative,xie2016temperature, xie2017variable}. It is very likely that the DMD technique presented in this paper could likewise be improved by incorporating these quantities as inputs to the forcing prediction algorithm, but that is left for future work.

The paper is structured as follows: Sec. \ref{sec:Background} presents an overview of DMD and its extension to time delay embedded coordinates. Sec. \ref{sec:dmd_forecasting} introduces DMD to construct linear control models directly from nonperiodic time series data, which can then be leveraged for probabilistic future-state prediction. Sec. \ref{sec:Results} presents numerical results for the application of this algorithm to real-world load data from the Duke Energy company, as well as a comparative analysis of other forecasting techniques applied to the same data. Finally, Sec. \ref{sec:Conclusion} offers concluding remarks and a brief discussion of possible future extensions for this work.

\section{Background: Dynamic Mode Decomposition}
\label{sec:Background}
The analysis developed here relies on the integration of two underlying mathematical architectures:  (i) the Dynamic Mode Decomposition, and (ii) Gaussian process regression.  Both methods are commonly used in a variety of data-driven applications where models are constructed directly from time-series measurements of a system of interest.

The forecasting method demonstrated in this work is based on the application of DMD to a time-delay representation of an observed data series~\cite{bruntonproctorkaiserkutz2017,arbabimezic2017,kamb18,dylewsky2020}. DMD is a regression method that produces a best-fit linear dynamical model to measured (generically nonlinear) time series data \cite{Schmid2010jfm,rowleymezicetal2009,Kutz2016book}.  It has been applied broadly in such diverse fields, for instance, as computer vision~\cite{grosek2014dynamic,erichson2019compressed}, fluid dynamics~\cite{Schmid2008aps,schmid2010,rowleymezicetal2009}, neuroscience~\cite{bruntonjohnsonojemannkutz2016}, finance~\cite{mann2016dynamic,hua2016using}, and epidemiology~\cite{proctor2015discovering}.

For some data matrix $\mathbf{X}$ whose columns are sequentially ordered snapshots of a system's state spaced evenly in time, DMD seeks a linear operator $\mathbf{A}$ which optimally satisfies
\begin{equation}
 \mathbf{x}^{(j+1)} = \mathbf{A}\mathbf{x}^{(j)}   
\end{equation}
where $\mathbf{x}^{(j)}=\mathbf{x}(t_j)$ for all $j$ (or, in the continuous time formulation, $\mathbf{\dot{X}} = \mathbf{A}\mathbf{X}$). In its simplest implementation, DMD is performed by solving the optimization
\begin{equation}
    \min_{\mathbf{A}} \|\dot{\mathbf{X}} - \mathbf{A}\mathbf{X}\|_F
\end{equation}
whose solution is $\mathbf{A} = \dot{\mathbf{X}}\mathbf{X}^\dagger$, 
where $\dagger$ denotes the Moore-Penrose pseudoinverse~\cite{turowleyetal2014} and the subscript $F$ denotes the Frobenius norm.
In this work an alternative formulation known as {\em optimized DMD} (optDMD) is used, where $\mathbf{A}$ is regressed by fitting exponential dynamics to the true observed signal using variable projection~\cite{askham2018variable}. In addition to improved signal reconstruction, this method 
can easily be constrained to seek out purely imaginary eigenvalues. This allows for the assumption that the measured dynamics occupy a post-transient regime. 
Obtaining a linear model for observed dynamics offers a number of practical advantages for understanding, controlling, and forecasting the system. $\mathbf{A}$ can be decomposed into its constituent eigenvectors and eigenvalues, which in turn can be used to write a closed-form solution to the DMD model. This representation offers a great deal of interpretability in terms of the coherent spatiotemporal structures which comprise the full signal~\cite{Kutz2016book}. 
Importantly, the DMD decomposition can be modified to handle streaming data~\cite{hemati2014dynamic,pendergrass2016streaming}, multiscale physics~\cite{kutz2016multiresolution,dylewsky2019}, and control~\cite{proctorbruntonkutz2016,proctor2018generalizing,mauroy2020introduction}.

\begin{figure}[t]
\centering
\includegraphics[width=0.76\columnwidth]{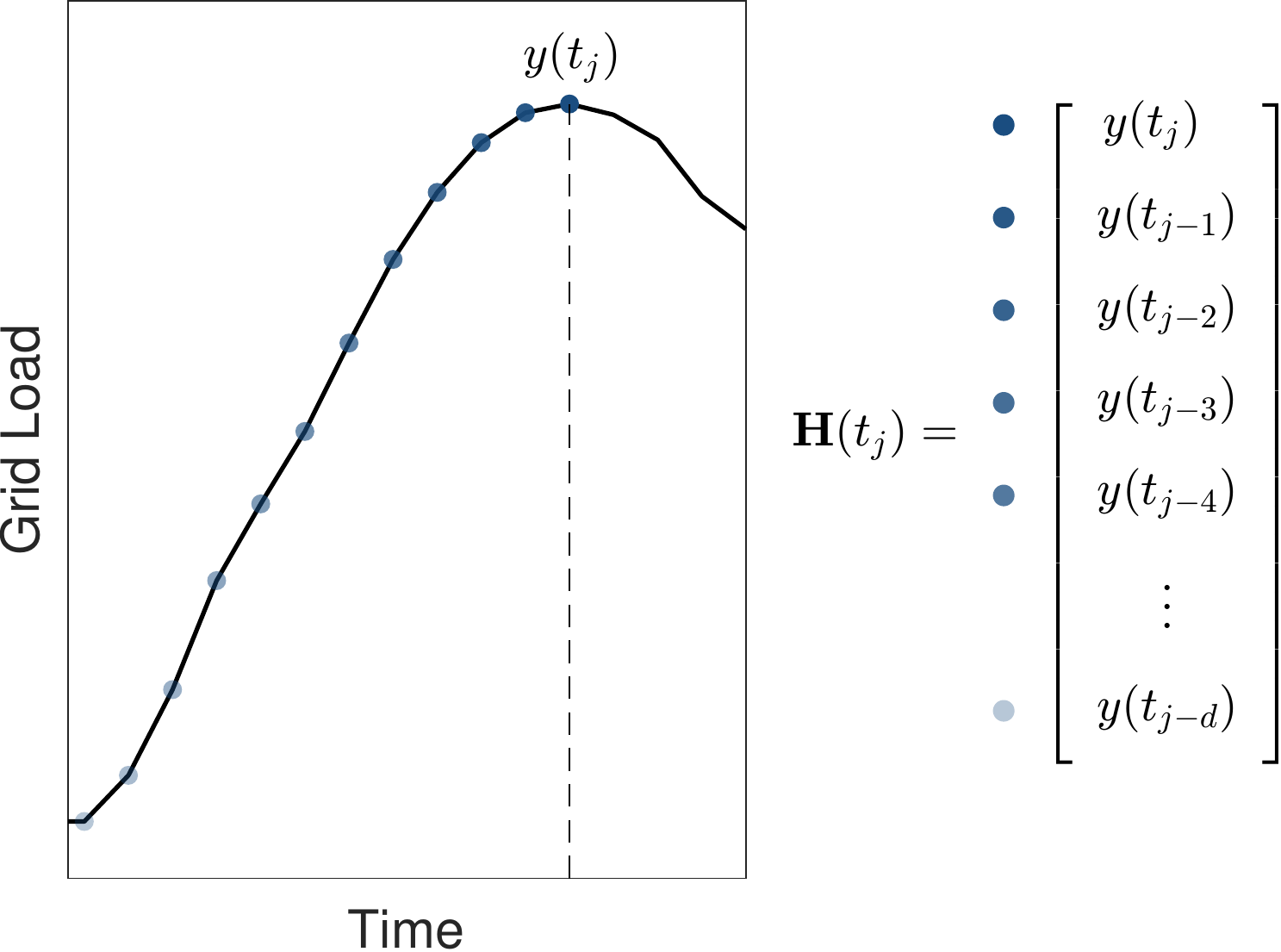}
\caption{Time delay embedding in which time-shifted copies of a scalar time series $y$ are stacked on top of one another to form a Hankel matrix $\mathbf{H}$. Each column of $\mathbf{H}$ can be thought of as a trajectory of $d$ sequential points along the trajectory in the original measurement space}
\label{fig:time_delay_embed}
\end{figure}

\begin{figure}[t]
\centering
\includegraphics[width=0.8\columnwidth]{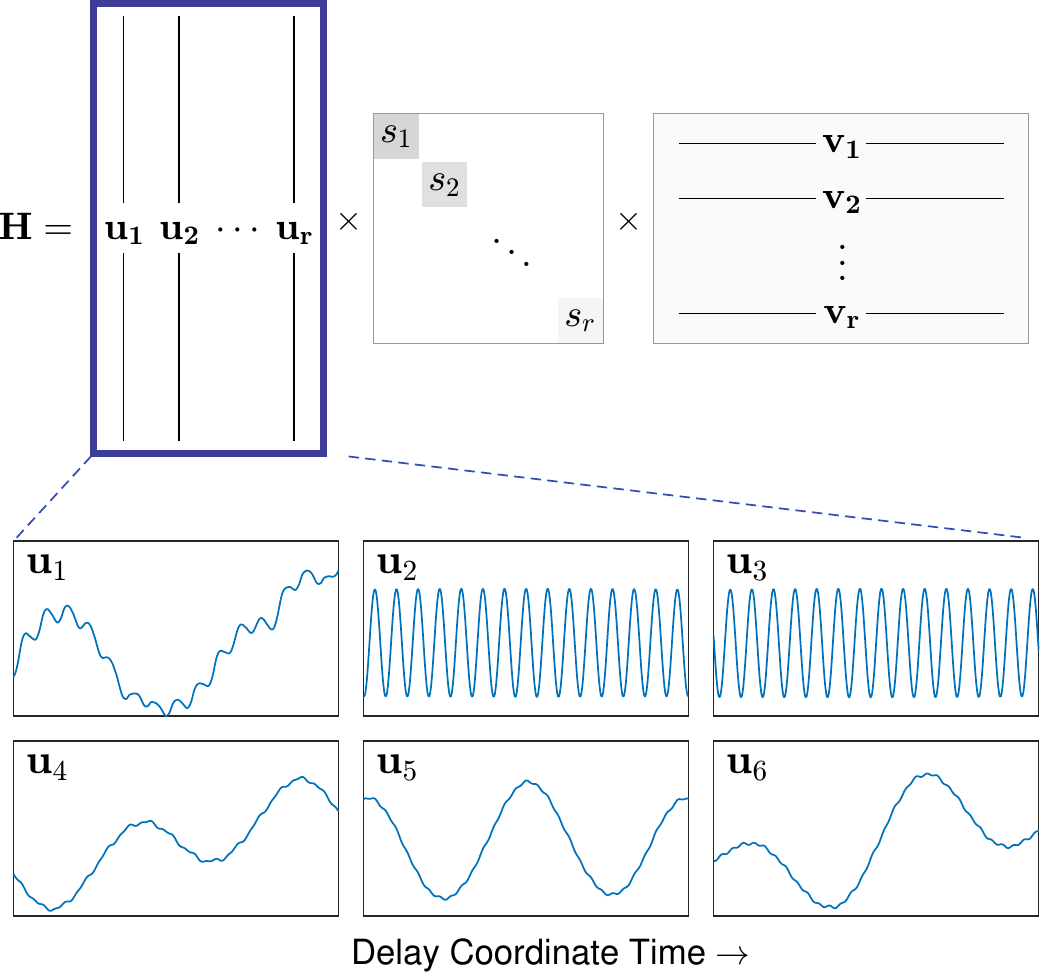}
\caption{Singular value decomposition of $\mathbf{H}$. Columns of $\mathbf{U}$ represent principal components in the delay coordinate space. These can also be represented as {\em principal component trajectories} (PCTs) in the state space, which can be thought of as a learned time-frequency basis for low-rank representation of the observed dynamics}
\label{fig:svd_modes}
\end{figure}

The rank of $\mathbf{A}$ cannot exceed the state dimension of $\mathbf{X}$, so for low-dimensional systems (such as scalar measurements of system-wide grid load), it is often useful to devise a transformation which `lifts' observation data into a higher-dimensional space in which the system might be more amenable to an accurate linear representation. This principle undergirds machine learning kernel methods such as support vector machines~\cite{scholkopf2002learning}, which relies on Cover's theorem~\cite{cover1965geometrical}. Many approaches to this lifting have been proposed, but one which has shown particular promise in recent years is time delay embedding~\cite{bruntonproctorkaiserkutz2017,arbabimezic2017,kamb18,dylewsky2020}. This method augments the state dimension simply by stacking time-shifted copies of a data series on top of each other to form a single delay matrix $\mathbf{H}$. As illustrated in Fig. \ref{fig:time_delay_embed}, columns of $\mathbf{H}$ which are state
vectors in these new delay coordinates can also be interpreted as trajectories of $d$ consecutive snapshots in the original state space, where $d$ is the number of delay embeddings used to construct $\mathbf{H}$. Application of DMD to trajectories lifted into this delay space (i.e. regressing $\mathbf{A}$ such that $\mathbf{H}^{(j+1)} \approx \mathbf{A}\mathbf{H}^{(j)}$ for all $j$) has been shown to be effective, as delay embedding guarantees that well-behaved dynamics appear increasingly linear as $d$ is increased \cite{bruntonproctorkaiserkutz2017,arbabimezic2017,panduraisamy2019}.
Indeed, the delay embeddings can produce so-called {\em principal component trajectories} (PCTs)~\cite{dylewsky2020}, which are the equivalent of ``principal component analysis'' modes for dominant low-rank trajectories.
This method bears some resemblance to {\em Singular Spectrum Analysis} (SSA)~\cite{bozzo2010relationship}, in which a scalar time series is analyzed by constructing a time delay matrix $\mathbf{H}$ and obtaining the singular value decomposition $\mathbf{H} = \mathbf{U}\mathbf{S}\mathbf{V}^T$ \cite{broomhead1989}. In this case the left singular vectors act as a learned basis for time-frequency analysis: recall that vectors in the time-delay space can also be interpreted as trajectories in the original state space, so each column of $\mathbf{U}$ represents a short dynamical template onto which local temporal dynamics can be projected (see Fig. \ref{fig:svd_modes}). 
Unlike DMD, however, this method does not provide a dynamical model or an interpretable eigenspectrum for the dynamics. However, for an infinite time-delay embedding, Bozzo et al.~\cite{bozzo2010relationship} show that SSA results in a perfect linear DMD model since the temporal modes correspond to pure sinusoids (exponents with purely imaginary eigenvalues).

\section{Forecasting with Time-Delay DMD}
\label{sec:dmd_forecasting}
\subsection{Methodology}

\begin{figure}[t]
\centering
\includegraphics[width=0.8\columnwidth]{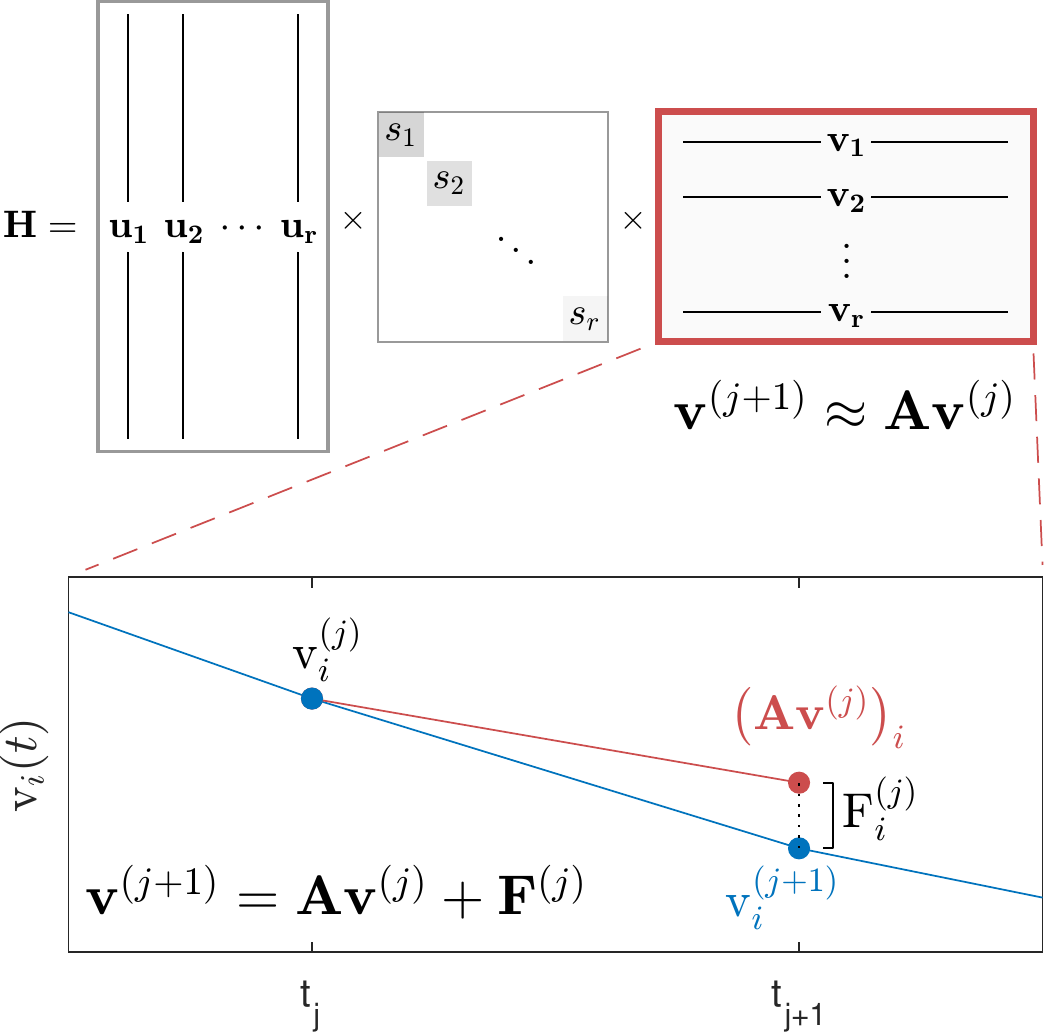}
\caption{The $\mathbf{V}$ matrix encodes the system's time series dynamics projected onto the top $r$ PCTs. Top: DMD regresses a best-fit linear operator $\mathbf{A}$ for the system's evolution in this basis. Bottom: this linear DMD model is used to make stepwise predictions of the state, and the discrepancy between these forecasts and the ground truth is interpreted as an external forcing $\mathbf{F}$ on the linear system}
\label{fig:dmd_protocol}
\end{figure}

The technique deployed in this work, which builds on the result of \cite{dylewsky2020}, combines the perspectives of DMD and SSA by building a linear model in a learned time-delay basis representation. A high-dimensional $\mathbf{H}$ is constructed by stacking $d$ time-shifted copies of the load time series $y(t)$ (large-$d$ limit), and then the dimension is reduced again by taking the SVD truncated at some rank $r << d$. This process of lifting and compressing allows for the signal to be projected into a low-rank time-frequency representation with fidelity which is optimal in the least-squares sense. A DMD model is then obtained in this basis via regression on the right singular vector matrix: $\mathbf{V}(t_{j+1}) \approx \mathbf{A}\mathbf{V}(t_j) ~\forall j$ (Fig. \ref{fig:dmd_protocol}, top).

This formulation is quite effective in producing approximate linear models for even highly nonlinear dynamics provided that those dynamics have a discrete Fourier spectrum. An $r \times r$ DMD operator $\mathbf{A}$ has at most $r$ distinct eigenvalues, which can map on to up to $r$ frequency peaks. Nonlinear systems with continuous Fourier spectra are therefore necessarily out of reach for DMD as no finite-dimensional linear operator can reproduce this phenomenon.
Nonetheless, there are many systems which are "almost periodic" in the sense that they exhibit Fourier spectra which are highly peaked but finite-valued in the gaps between the peaks. For these cases, the delay-coordinate DMD method can be extended to learn an \textit{actuated} linear model \cite{dylewsky2020}: After solving for the DMD operator $\mathbf{A}$, stepwise forecasts are made for each observed data point: $\tilde{\mathbf{v}}^{(j+1)} = \mathbf{A}\mathbf{v}^{(j)}$. Comparing these predictions to the true $\mathbf{v}^{(j+1)}$, the residual can be labeled as a parametric forcing $\mathbf{F}(t)$ on the linear model: $\mathbf{F}^{(j)} = \mathbf{v}^{(j+1)} - \tilde{\mathbf{v}}^{(j+1)}$ (Fig. \ref{fig:dmd_protocol}, bottom). The full actuated system with dynamics defined by $\mathbf{v}^{(j+1)} = \mathbf{A}\mathbf{v}^{(j)} + \mathbf{F}^{(j)}$ then perfectly reproduces the original signal by construction.

The proposed modeling approach offers a representation in which the dynamics can be understood as a simple linear system with (quasi-)periodic intrinsic solutions being forced by a nonperiodic external actuator. The former component encompasses the portion of the dynamics which is recurring and therefore able to be deterministically extrapolated into the future. The latter contains a distillation of the dynamical content which does not meet this description. Previous work on Koopman representations for systems with continuous spectra has made use of the result that measure-preserving dynamical evolution in the post-transient regime shares a duality with stationary stochastic processes \cite{arbabimezic2017prf}. For the problem of forecasting in almost-periodic systems (i.e. those with a continuous spectrum interspersed with dominant narrow peaks accounting for a large fraction of the total energetics), this stochasticity can be relegated to the forcing signal described above. 

Forecasting can then be accomplished by fitting a suitable stochastic model to the forcing obtained over the training period. Evolution into the future is simulated by integrating a linear control model forward in time and continuously sampling the stochastic process to supply the actuation. As this paper will show, this approach outperforms many widely-used numerical forecasting methods when applied to a real-world data set of system-wide grid load. Moreover, it integrates naturally with uncertainty quantification methods: these forced linear systems are constructed such that all model error definitionally lives in the actuation signal, so one can quantify forecasting uncertainty by sampling an ensemble of realizations of the stochastic process that models it. Finally, this method offers a novel mode of interpretability in data-driven modeling: In addition to the spectral information granted by eigendecomposition of the obtained DMD operator, the learned forcing signal can be analyzed on its own, which offers insight into anomalous events in the data.

\subsection{Mathematical Formulation}
\label{sec:mathematical_formulation}
The only required input for this method is a time series $y(t)$ consisting of a number of measurements $m$ which is large compared to the desired delay embedding dimension $d$. For the purposes of this analysis, we further assume that $y(t)$ is almost-periodic in the sense described in the previous section, and approximately stationary.
The data set used in this work covers a duration of a full year and therefore does contain seasonal effects which could undermine the assumption of stationarity, but models are exclusively trained on 4-week subsamples over which the impact of seasonality is marginal.

The input data $\mathbf{y}$, expressed as a $1\times m$ vector, is lifted into delay coordinate space by defining the Hankel matrix $\mathbf{H}$:
\begin{equation}
\begin{split}
\mathbf{H} &= \left[\begin{matrix} y(t_1) & y(t_2) & \cdots & y(t_{m-d}) \\
									 y(t_2) & y(t_3) & \cdots &y(t_{m-d+1})\\
									 \vdots & \vdots & \ddots & \vdots\\
									 y(t_d) & y(t_{d+1}) & \cdots &y(t_m)\end{matrix}\right].
\end{split}
\label{eq:time_delay}
\end{equation}
$\mathbf{H}$ has dimension $d\times(m-d)$, where $d$ is the chosen number of delay embeddings. We choose $d$ such that the delay embedding duration $(t_d - t_1)$ is long enough to encompass the dominant periodicity of the data (in this case, daily and weekly oscillations), but short enough that $d<<m$. For this analysis we set $d = 362$, which corresponds to an embedding duration of just over 2 weeks. 

The delay-embedded data is then decomposed using SVD:
\begin{equation}
    \begin{split}
        \mathbf{H} &= \mathbf{U} \mathbf{S} \mathbf{V}^T.
    \end{split}
\end{equation}
At this point we apply a windowing process, isolating a 4-week subset of the time series matrix $\mathbf{V}$ (note that the time-dispersed nature of vectors in delay coordinates mean that this window actually contains information from the previous $4+(t_d-t_1) \approx 6$ weeks). This windowing process is repeated for 8 randomized subsets of the data, and the computations that follow are repeated for all of them.

The linear model for the delay-coordinate dynamics is obtained by applying DMD to each windowed time series from $\mathbf{V}$. For the sake of model simplicity, only the first $r=16$ columns of $\mathbf{V}$ are used (corresponding to the top 16 SVD modes):
\begin{equation}
    \begin{split}
        \left[\begin{matrix}
            | & | & & | \\
            \mathbf{v}^{(2)} & \mathbf{v}^{(3)} & \cdots & \mathbf{v}^{(w)}\\
            | & | & & | \\
        \end{matrix}\right]&
        \!\!\approx \!\!
                \mathbf{A}\!
        \left[\begin{matrix}
            | & | & & | \\
            \mathbf{v}^{(1)} & \mathbf{v}^{(2)} & \cdots & \mathbf{v}^{(w-1)}\\
            | & | & & | \\
        \end{matrix}\right].
    \end{split}
\end{equation}
The linear operator $\mathbf{A}$ is regressed using the optDMD algorithm constrained to admit only imaginary eigenvalues (i.e., stable oscillation, without growth or decay.)

The forcing is computed by performing a single forward time step and differencing the result with the ground truth:
\begin{equation}
    \begin{split}
        \mathbf{F} =& \mathbf{V} - \mathbf{\tilde{V}} = \left[\begin{matrix}
            | & | & & | \\
            \mathbf{v}^{(2)} & \mathbf{v}^{(3)} & \cdots & \mathbf{v}^{(w)}\\
            | & | & & | \\
        \end{matrix}\right]
        -\\
        &\left[\begin{matrix}
              & & & & \\
              & &\mathbf{A}& & \\
              & & & & \\
        \end{matrix}\right]
        \left[\begin{matrix}
            | & | & & | \\
            \mathbf{v}^{(1)} & \mathbf{v}^{(2)} & \cdots & \mathbf{v}^{(w-1)}\\
            | & | & & | \\
        \end{matrix}\right].
    \end{split}
\end{equation}
The forced linear model defined by $\mathbf{v}^{(j+1)} = \mathbf{A}\mathbf{v}^{(j)} + \mathbf{F}^{(j)}$ then by construction fits the observed data perfectly. It should be noted that while the DMD model was trained on rank truncated SVD data, $\mathbf{F}$ is computed in the full delay embedding dimension $d$. All columns of $\mathbf{V}$ above $r=16$ are treated as residual to the linear model and passed directly into $\mathbf{F}$. This choice reflects the assumption that the (quasi-)periodic linear dynamics are spectrally sparse, and can be well captured by $r$ DMD eigenvalues. The nonperiodic contribution to dynamics, however, is expected to have a long spectral tail and therefore require a much higher-dimensional representation.

To forecast with this model, an auxiliary stochastic model must be obtained to estimate values for $\mathbf{F}$ beyond the training window.
In this work we employ a GPR approach described as follows: A GP with a Mat{\'e}rn kernel (and shape parameter $\nu = 3/2$) is trained to take as input a vector containing both the current delay-coordinate state $\mathbf{H}(t_j)$ and a recent history of forcing values $\mathbf{F}(t_{j-q}),\mathbf{F}(t_{j-q+1}),...,\mathbf{F}(t_{j-1})$.
We choose $q=16$, but this selection could likely be refined via model selection.
The forcing $\mathbf{F}$ is a vector in $\mathbb{R}^d$, which necessitates $d$ separate GPR models (each providing only a single output.)
At $d=362$ this would be highly demanding, so $\mathbf{F}$ is first projected onto its top $r_F = 8$ SVD modes.
An alternative approach is to employ multiple-output/multivariate GPR; in fact, one might expect that estimating $\mathbf{F}(t_j)$ using 8 independent Gaussian processes might be less effective than a method which accounts for correlations among the outputs, but empirical experimentation suggested that building in such a correlating influence offered negligible improvement to accuracy. 

With this auxiliary stochastic model in place, forecasting can be carried out by repeated computation of $\mathbf{v}^{(j+1)} = \mathbf{A}\mathbf{v}^{(j)} + \mathbf{F}^{(j)}$, with $\mathbf{F}^{(j)}$ being a sample of the trained Gaussian process.
This can be repeated as many times as desired to build up an ensemble of stochastic realizations for the forecast.

\section{Results}
\label{sec:Results}
\subsection{Unforced DMD in delay coordinates}
\begin{figure}[t]
\centering
\includegraphics[width=0.8\columnwidth]{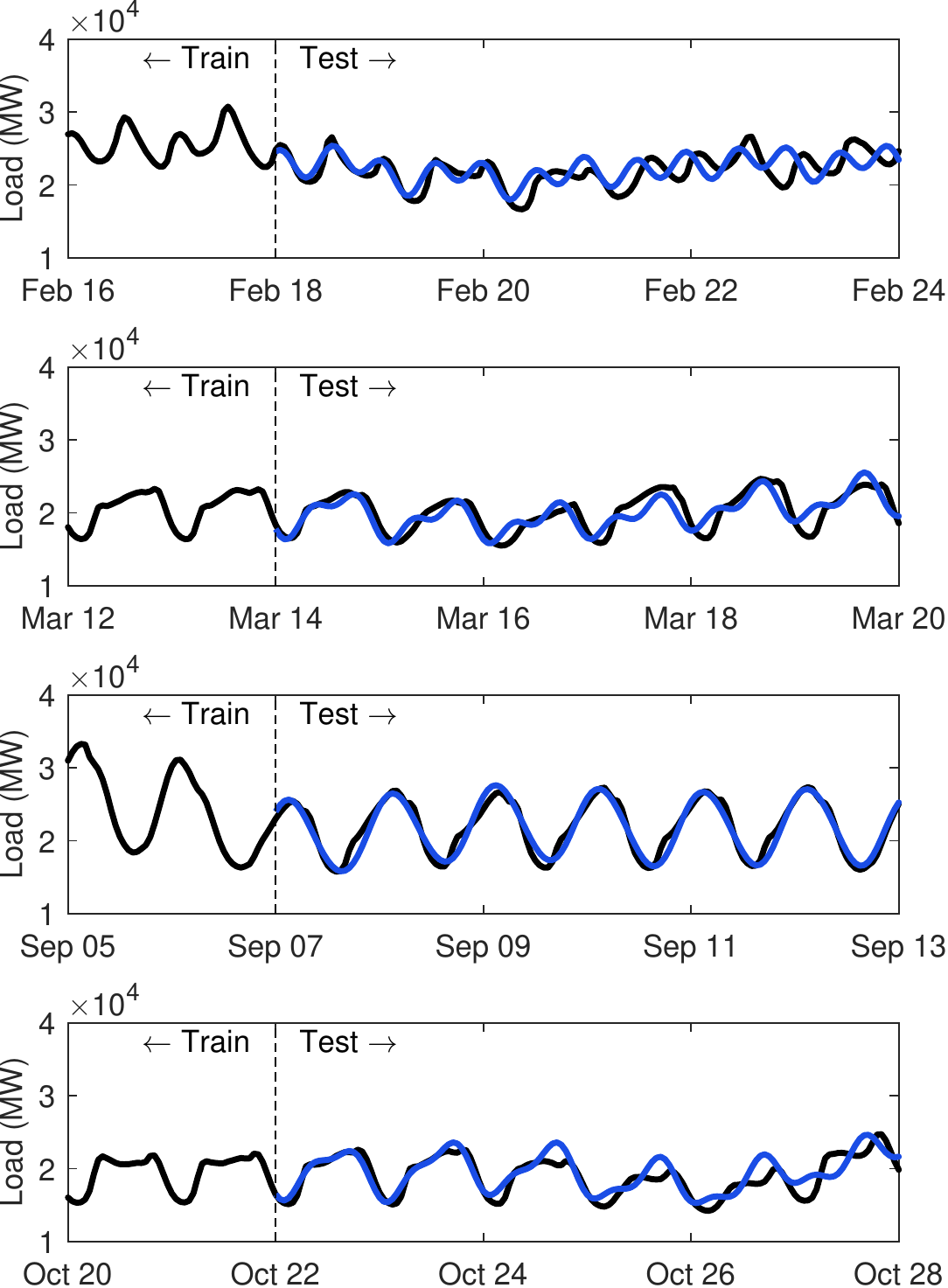}
\caption{Unforced DMD forecasts for 4 windowed subsets of load data. Models trained on a 4 week history are integrated forward in time approximately 1 week (blue) and plotted against true load values over that time (black)}
\label{fig:dmd_forecast_unforced}
\end{figure}

Forecasting results for the learned delay-coordinate DMD models \textit{without} forcing are plotted in Fig. \ref{fig:dmd_forecast_unforced}. These models were trained on a 4 week window (not plotted in full) and integrated out approximately 1 week into the future. In many cases prediction accuracy is quite good over this duration, though phase errors do often begin to creep in near the end.

\begin{figure}[t]
\centering
\includegraphics[width=0.8\columnwidth]{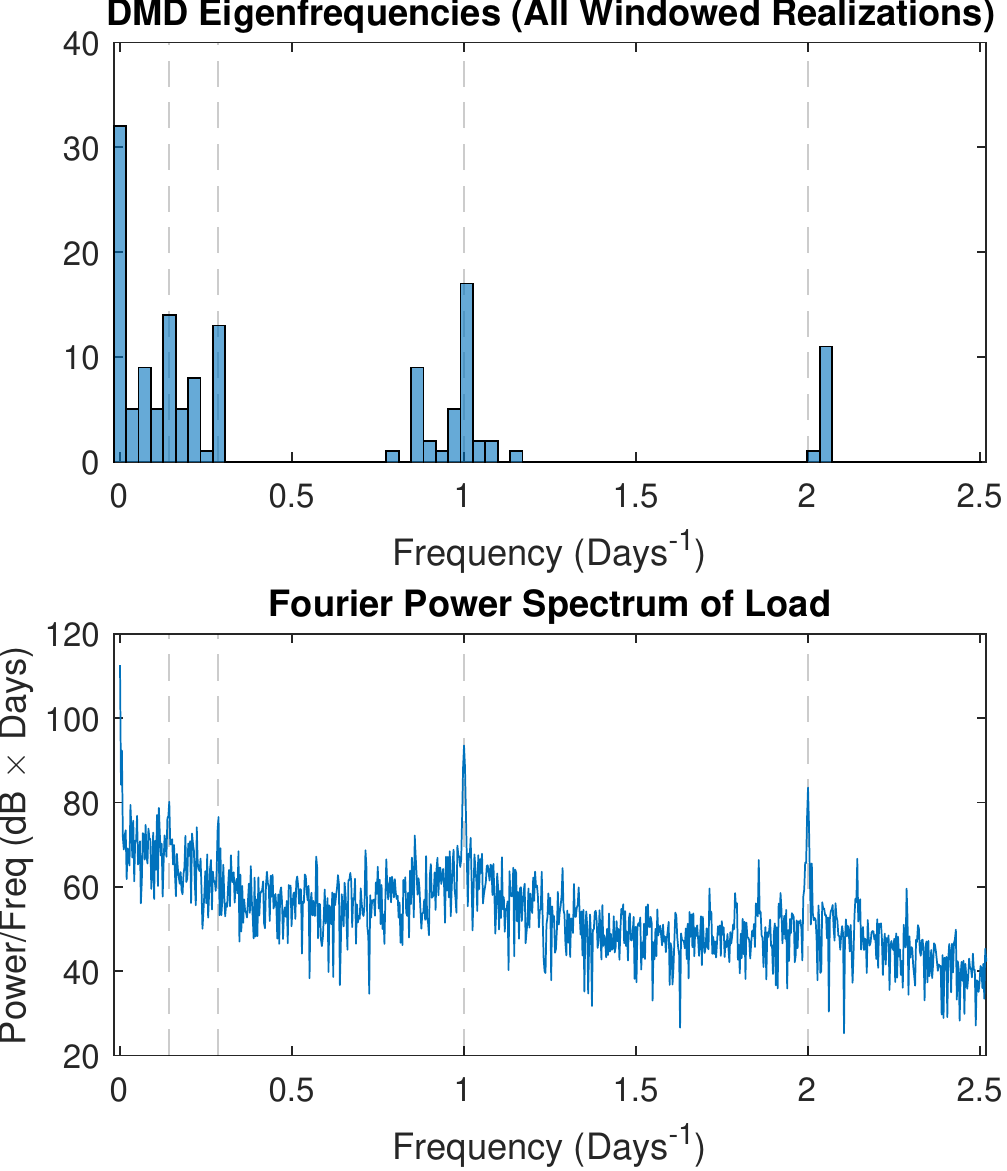}
\caption{Top: Histogram of eigenvalues from models trained on randomized 4-week subsets of the Duke load data. Note that while linear models generically allow for complex eigenvalues, the DMD algorithm used here restricts eigenvalues to the imaginary axis. Bottom: Fourier power spectrum of the full (1 year) load data. Dotted lines on both plots denote dominant frequency components with periods of 7, 3.5, 1, and 0.5 days.}
\label{fig:dmd_spectrum_analysis}
\end{figure}

One advantage of the linear models produced by DMD is that their dynamics can be understood in terms of eigenvalues and eigenvectors of the matrix $\mathbf{A}$. As discussed in Section \ref{sec:dmd_forecasting}, the expectation of these models is that they capture dominant peaks in the power spectrum. Eigenvalues from models from all 4 week training windows are plotted in a histogram in Fig. \ref{fig:dmd_spectrum_analysis}. The frequencies most commonly represented in the DMD models clearly correspond to peaks present in the Fourier power spectrum plotted below. The spectral bands between peaks have little to no representation in the DMD eigenspectra, so for a linear control model to obtain perfect fidelity to observed dynamics those frequencies will need to be stimulated by external actuation.

\subsection{Forced linear models}
\begin{figure}[t]
\centering
\includegraphics[width=0.88\columnwidth]{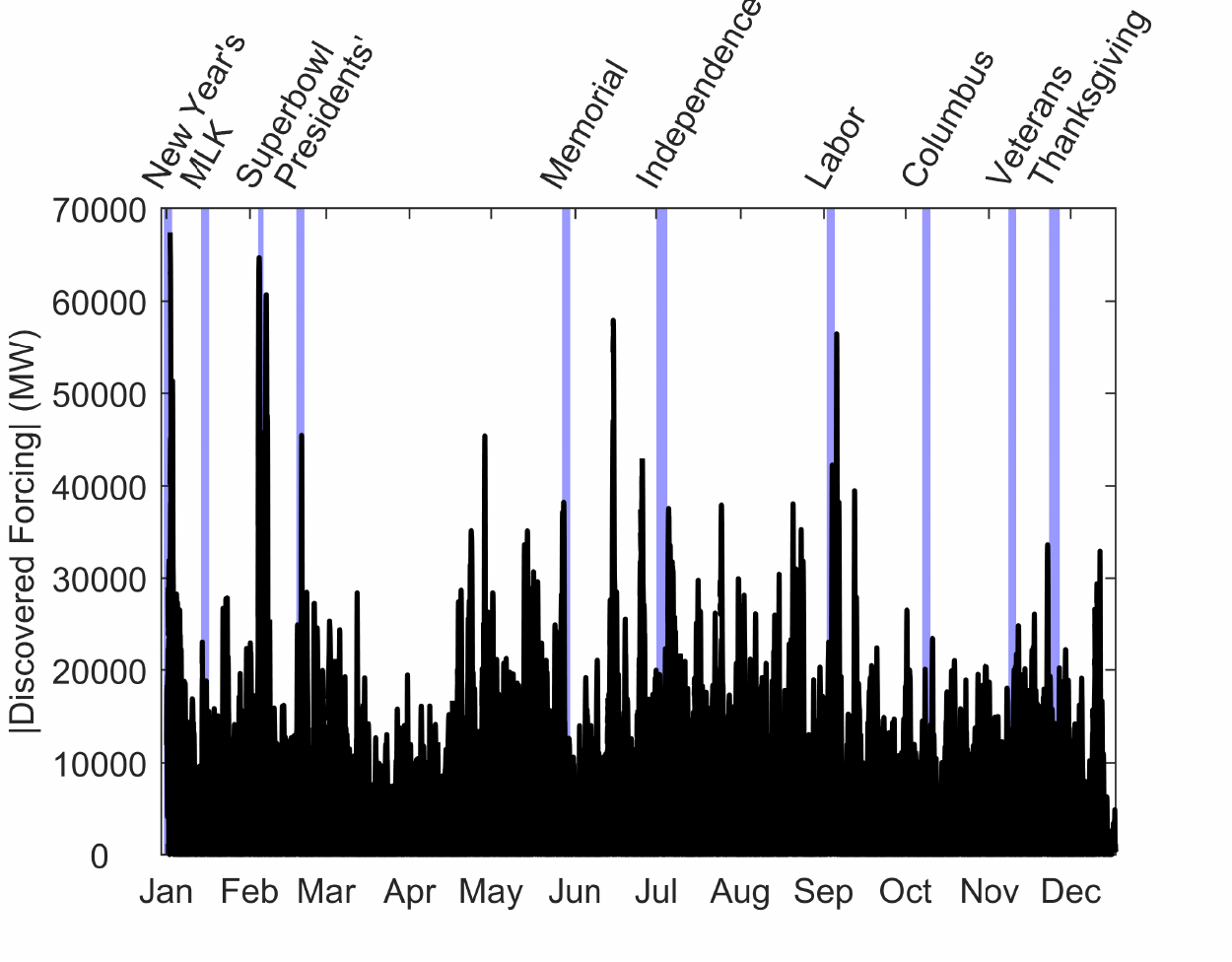}
\caption{Absolute value of discovered forcings for the Duke load data. Model regression and forcing inference are carried out on overlapping 4 week windows, results for which are overlaid in this figure.}
\label{fig:forcing_interpretation}
\end{figure}

As outlined in Section \ref{sec:mathematical_formulation}, a forcing signal for the specified training window is learned in an unsupervised fashion by collecting residual errors of stepwise linear forecasts into a single time series. Although the resulting linear control model simply reproduces the dynamics on which it was trained (by construction), it does so in a manner which separates linearizable evolution (i.e. behavior with discrete Fourier spectrum) from the rest of the (spectrally continuous) dynamics. The latter component, which is relegated to the learned forcing, can offer valuable diagnostic information into the system in question.
Grid load is driven by complex environmental factors including climate, economic, and social variables, all of which can be considered nonperiodic actuators on whatever natural cyclic behaviors (daily, weekly, etc.) dominate the observed dynamics.
A learned forcing signal in this case would certainly include contributions from all of these factors, but not in any separable or easily interpreted form. Nonetheless, certain qualitative observations can be offered to support the interpretation that the obtained forcing corresponds to anomalies in power consumption. Results for the (absolute value of) the forcing signal obtained from applying this method to the Duke load data are plotted in Fig.~\ref{fig:forcing_interpretation}.
Highlighted in blue are a number of likely candidates for days of unusual human behavior with respect to electricity usage, including all federal holiday weekends and the Superbowl.
Visual inspection suggests a clear correspondence between these periods and large spikes in forcing. Many additional spikes are observed which do not fall on any of these days, but these could be accounted for by any number of other possible disturbances: locally observed holidays, temperature anomalies such as heat waves and cold snaps, major world events, etc.

\subsection{Forecasting}

\begin{figure}[t]
\centering
\includegraphics[width=0.86\columnwidth]{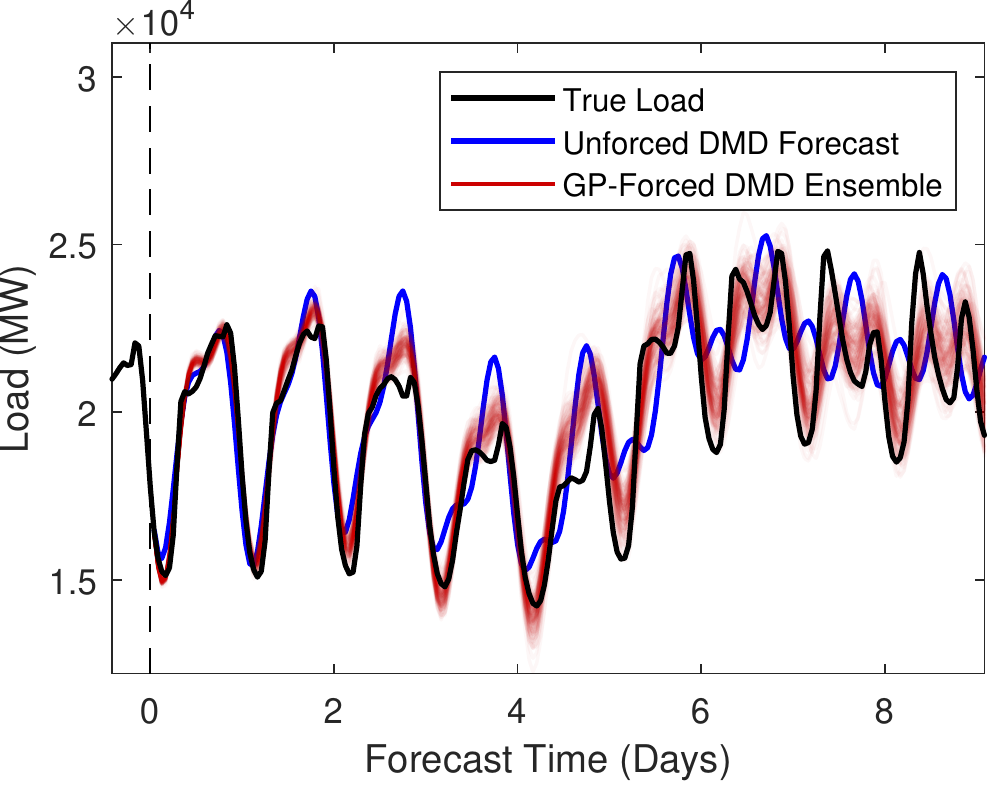}
\caption{Future state predictions of a delay-coordinate DMD model without forcing (blue) and with forcing stochastically sampled from a Gaussian process trained on the preceding 4 weeks (red), plotted against the observed ground truth (black). The dotted vertical line denotes the boundary between the 4 week training window (mostly cut off) and the test domain.}
\label{fig:gpr_dmd_ensemble_forecast}
\end{figure}

Stochastic ensemble forecasting is carried out as described in Sec. \ref{sec:mathematical_formulation}. Fig. \ref{fig:gpr_dmd_ensemble_forecast} illustrates an example of results for a linear control model built on a 4 week training window and run forward approximately a week into the future. 128 realizations of the Gaussian process randomized forcing are sampled to generate a density cloud of possible future trajectories. It can be seen that forced trajectories (red) more closely match the true signal (black) than the unforced DMD forecast (blue). In particular, the phase error that the latter accumulates over time is corrected by the forcing.

For the purpose of validation, 2 week forecasts were carried out for all 4 week test windows out using a number of data-driven methods which do not rely on external explanatory variables.
We compare the forced and unforced variations of the proposed DMD forecasting method against predictions computed using long short-term memory (LSTM) recurrent neural networks, ARIMA, GPR, and ensemble GPR~\cite{Tong_EGPR}.
All methods were supplied with the same 4 week training sets.
The LSTM network consisted of a single LSTM layer with 200 hidden units ($\tanh$ activation) connected to a scalar output layer.
The ARIMA model used parameters $p=24$, $d=1$, and $q=1$, chosen using the Box-Jenkins method of model selection. 

The GPR method used for comparison differs from the one previously used for modeling the forcing on the DMD models.
Specifically, here we employ a traditional approach for time series forecasting, which consists of assuming that the time series to be modeled is a realization of a GP with prescribed parameterized mean and covariance.
The hyperparameters of the mean and covariance are estimated by maximizing the marginal likelihood of the data~\cite{williams2006gaussian}.
In this setting, GPR is a univariate regression model that does not rely on explanatory variables for load forecasting~\cite{yang2018power,shepero2018residential,lloyd2014gefcom2012}.
We employ the following parameterized periodic covariance function:
\begin{equation}\label{sgpr_co}
  \ K(t, \tau) = 
   \gamma^2_1 \exp \left \{ - \frac{2 \sin^2 \left [ \frac{\pi}{24} (t - \tau) \right ]}{\gamma^2_2} \right \},
\end{equation}
with hyperparameters $\gamma_1$ and $\gamma_2$.

Results are also presented for the Ensemble Gaussian Process Regression (EGPR) method, which has been shown to improve accuracy in load forecasting applications compared to the standard GPR approach described above~\cite{Tong_EGPR}.
As in standard GPR, data is modeled as the realization of a GP with certain mean and covariance.
Instead of employing parametric models for these quantities, in EGPR they are estimated from sample statistics of windowed snapshots of the data.
To ameliorate the rank deficiency of the sample covariance matrix when the number of snapshots is low, we regularize this sample estimator using Rao-Blackwell-Ledoit-Wolf shrinkage~\cite{chen2010shrinkage}.

\begin{figure}[t]
\centering
\includegraphics[width=0.9\columnwidth]{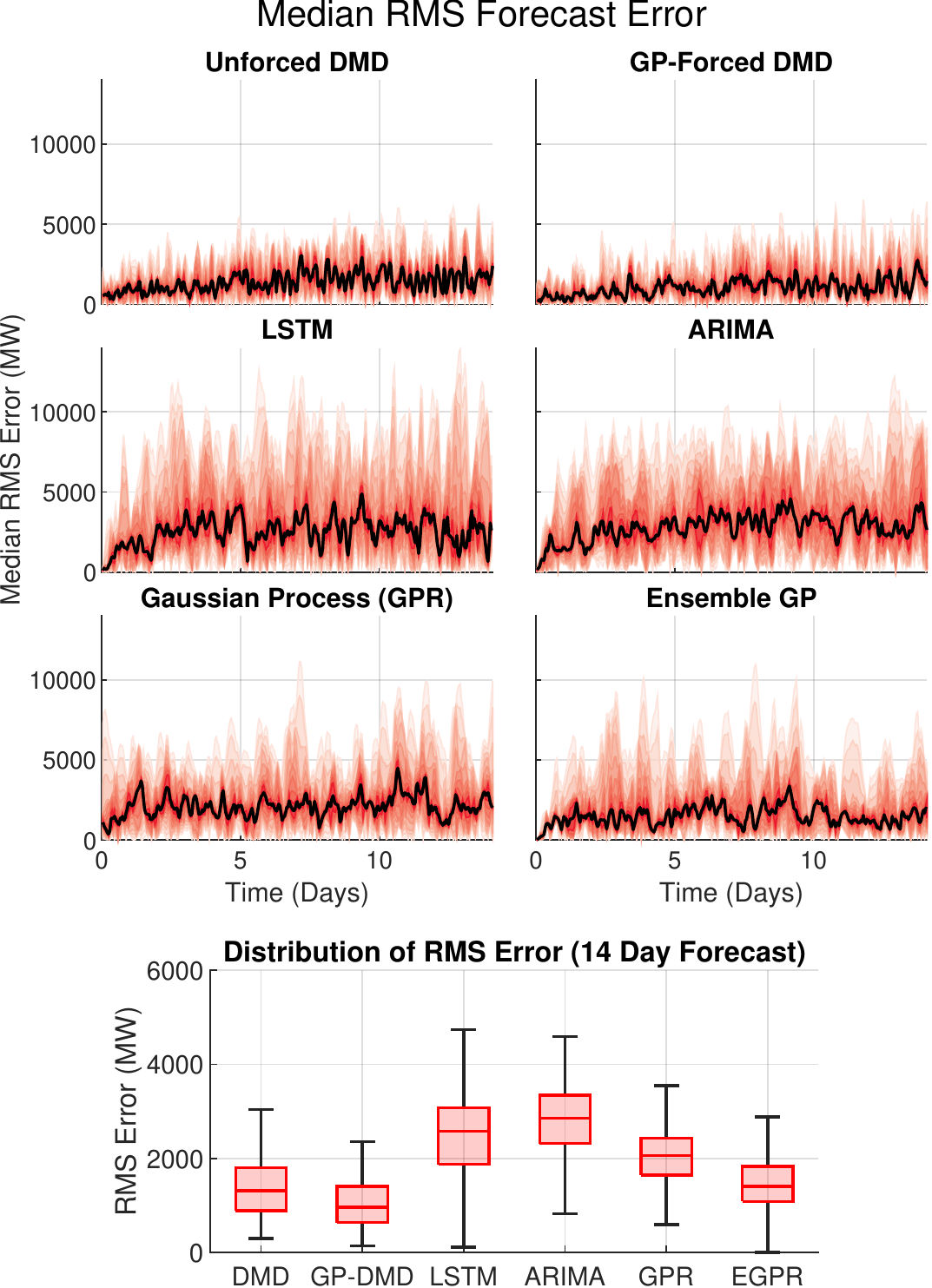}
\caption{Comparison of forecasting error for 6 different models. Upper plots show evolution of the RMS prediction error over 2 weeks. The black lines represent medians for each model over all realizations on different 4 week training windows. The surrounding red shading gives distribution statistics for the same data (each successively lighter band denotes another 5 percentile). The lower plot shows statistical distributions of these RMS errors averaged over the 14 day prediction window.}
\label{fig:compare_forecast_methods}
\end{figure}

A statistical comparison between all of these forecasting methods over all 4 week training windows is given in Fig. \ref{fig:compare_forecast_methods}. The upper plots show median RMS forecast error of each method out to 2 weeks in the future. The lower plot depicts statistics of these errors averaged over the forecast window. The forced DMD model offers the best performance of all the methods, and even the unforced model is competitive with the other leading contenders.

\section{Conclusion}
\label{sec:Conclusion}

The method proposed in this work offers a purely data-driven means for constructing linear control models to reconstruct and forecast time series with almost-periodic dynamics.
We apply the proposed method to power grid load forecasting, in which predictably oscillatory behavior is dominant but nonetheless subject to perturbations resulting from coupling to environmental variables including weather patterns, economic trends, and variations in human behavior. As a forecasting method, the success of this technique lies in its ability to model observation data in a manner that separates these two aspects of its governing dynamics. Components of the dynamics which cannot be feasibly modeled (i.e. effects due to noise and/or chaotic external forcing) are modeled statistically, but this stochasticity is kept separate from the dominant quasiperiodic dynamics. We have shown that the forecasting performance of this method on a data set of real-world grid load is superior to other leading data-driven methods (those which do not make use of explanatory variables). Moreover, this algorithm offers greater interpretability as a forced linear system whose learned actuation can offer insight into anomalous events in the system's environment.

In future work, we hope to see improved understanding of how the learned forcing signal can be interpreted through its correlation to real-world events. We have shown that large spikes in the forcing can often be attributed to holidays which disrupt usual power consumption patterns, but a more sophisticated analysis might elucidate more subtle correlations with temperature and precipitation data, electricity prices, and other variables known or suspected to influence power consumption. A more thorough account of the relationships could be used to improve the stochastic modeling of the control signal by using these quantities as input features.

Additionally, further exploration of how forcing is modeled could allow for greater user control over uncertainty in ensemble forecasting. In this work we have constructed systems forced by a kick at every time step, but this could be reconfigured so that a proportionately stronger actuation is applied at sparser intervals. If a future state prediction is computed by sampling a stochastic process each time a new kick is applied, a smaller number of such samples would translate to a larger variance in the resulting forecast ensemble. This could even be tuned using training data to obtain a propagation of forecast uncertainty which better matches observed dynamics. Although we have focused our discussion in this paper on power grid load forecasting, this method should generalize well to many complex, noisy, or stochastic systems.

\section*{Acknowledgment}

This work was supported by the U.S. Department of Energy, Office of Science, Office of Advanced Scientific Computing Research (ASCR) under Contract DE-AC02-06CH11347.
Pacific Northwest National Laboratory is operated by Battelle for the DOE under Contract DE-AC05-76RL01830.

\bibliographystyle{ieeetr}
\bibliography{main}   

\end{document}